\documentclass[12pt]{article}

\usepackage[dvipdfmx]{graphicx}
\usepackage[dvipdfmx]{xcolor}
\usepackage{amsmath}
\usepackage{amssymb}
\usepackage{tikz}
\usepackage{cite}
\usepackage{fancyhdr}

\def\erase#1{{}}
\def\EqArrerase#1{{}}

\makeatletter
 
 \@addtoreset{equation}{section}
\makeatother

\def\GL{{G\kern-.12em L\kern-.04em}}
\def\OSp{{O\kern-.11em S\kern-.04em p}}
\def\IOSp{{I\kern-.06em O\kern-.11em S\kern-.04em p}}
\def\MN{{M\kern-.14em N}}
\def\NM{{N\kern-.14em M}}
\def\NL{{N\kern-.14em L}}
\def\LN{{L\kern-.11em N}}
\def\ML{{M\kern-.14em L}}
\def\LM{{L\kern-.11em M}}
\def\RN{{R\kern-.11em N}}
\def\NR{{N\kern-.14em R}}
\def\RM{{R\kern-.11em M}}
\def\MR{{M\kern-.14em R}}
\def\RL{{R\kern-.11em L}}
\def\LR{{L\kern-.11em R}}
\def\RS{{R\kern-.11em S}}
\def\SR{{S\kern-.11em R}}
\def\SN{{S\kern-.11em N}}
\def\NS{{N\kern-.11em S}}
\def\SM{{S\kern-.11em M}}
\def\MS{{M\kern-.11em S}}
\def\SL{{S\kern-.11em L}}
\def\LS{{L\kern-.11em S}}
\def\sqr#1#2{{\vcenter{\hrule height.#2pt
      \hbox{\vrule width.#2pt height#1pt \kern#1pt
          \vrule width.#2pt}
      \hrule height.#2pt}}}
\def\bra0{\langle0|}
\def\ket0{|0\rangle}
\def\soeji#1_#2#3{#1_{#2}\cdots#1_{#3}}
\def\longgLRarrow{\longleftarrow\kern-3pt\relbar\kern-3pt\relbar\kern-3pt%
\longrightarrow}
\def\longLRarrow{\longleftarrow\kern-3pt\relbar\kern-3pt\longrightarrow}
\def\longLarrow{\longleftarrow\kern-3pt\relbar\kern-3pt\relbar\kern-3pt\relbar}
\def\longRarrow{\relbar\kern-3pt\relbar\kern-3pt\relbar\kern-3pt\longrightarrow}
\def\bothDer#1#2#3{%
\overset{\kern-.7em\stackrel{#1}{#2}}{\partial_{#3}}}
 
\makeatletter
 
 \@addtoreset{equation}{section}
\makeatother

\begin{document}
\thispagestyle{fancy}

\title{Confinement of Massive Ghost in Quadratic Gravity}

\author{Ichiro Oda
\footnote{Electronic address: ioda@cs.u-ryukyu.ac.jp}
\\
{\it\small
\begin{tabular}{c}
Department of Physics, Faculty of Science, University of the 
           Ryukyus,\\
           Nishihara, Okinawa 903-0213, Japan\\      
\end{tabular}
}
}
\date{}

\maketitle

\thispagestyle{fancy}

\begin{abstract}

In the framework of the covariant canonical formalism of quadratic gravity, we consider the problem of confinement of massive ghost which
violates the unitarity of the physical S-matrix. It is shown that  if there is a bound state between the massive ghost and Faddeev-Popov ghost
the massive ghost is confined in the zero-norm states through the BRST quartet mechanism, thereby the unitarity being restored. Based on 
the superfield formulation by Bonora and Tonin, we show that the asymptotic field of the massive ghost must be a massive dipole 
whereas that of the bound state obeys a massive Klein-Gordon equation. This situation may be of some similarity to color confinement 
in quantum chromodynamics (QCD) where it is conjectured that not a massless but a massive gluon is in fact confined.

\end{abstract}

\newpage
\pagestyle{plain}
\pagenumbering{arabic}


\section{Introduction}

One of the most interesting attempts to modify Einstein's general relativity is to add to the Einstein-Hilbert action quadratic terms in scalar curvature 
and conformal tensor, whose theory has recently been called ${\it{quadratic \, gravity}}$, since such the terms naturally appear as quantum effects 
in the semiclassical theory where the gravitational field is treated classically whereas the matter fields are treated quantum mechanically \cite{BD}.
It is worthwhile to stress that quadratic gravity is not only perturbatively renormalizable  \cite{Stelle} but also its dimensionless coupling constants
are asymptotically free \cite{Mario, Fradkin, Avramidi}.\footnote{Speaking more precisely, the coupling constant in front of the $R^2$ term might
not be asymptotically free \cite{Savio}.} 

These properties make quadratic gravity very attractive as a  promising candidate of quantum gravity, but the price we have to pay is very high.
In the linearized theory around a flat Minkowski background (in case of the vanishing cosmological constant), the spectrum contains a massive
spin-2 ghost called ${\it{massive \, ghost}}$ in this article, in addition to the massless spin-2 graviton as well as the massive scalar sometimes called 
${\it{scalaron}}$ in cosmology or ${\it{graviscalar}}$ in particle physics. Since the massive ghost violates the unitarity 
of the physical S-matrix, quadratic gravity has never been accepted as a viable solution to the problem of quantum gravity for a long time.
Even if a lot of ideas towards the resolution to the problem of the massive ghost have been proposed thus far, no one has succeeded in
convincing our communitity completely. 

The answer to this problem must present a mechanism to get rid of the massive ghost from the physical Hilbert space.  Such an answer 
has previously been proposed in Refs. \cite{Kawasaki, Kimura2} where the massive ghost is confined in the zero-norm states through 
the BRST quartet mechanism \cite{Kugo-Ojima, Nishijima} as in the Faddeev-Popov (FP) ghosts in the Yang-Mills theory. One of motivations in this
article is to reconsider this answer in the covariant canonical formalism of quadratic gravity \cite{Oda-Can} and shed different light on it by using the 
superfield formalism based on six-dimensional superspace $(x^\mu, \theta, \bar \theta)$ by Bonora and Tonin \cite{Bonora-Tonin}.  

Here,  from our viewpoint let us briefly review the problem of the unitarity violation in quadratic gravity owing to the massive ghost. The action 
of quadratic gravity is of form\footnote{The detail will be explained later.}
\begin{eqnarray}
S = \int d^4 x \, \sqrt{-g} \left( \frac{1}{2 \kappa^2} R - \alpha_C C_{\mu\nu\rho\sigma}^2 + \alpha_R R^2 \right).
\label{QG-action-0}  
\end{eqnarray}
Let us first recall where the massive ghost stems from this action. Taking quadratic terms in $\varphi_{\mu\nu}$ from the first and second terms 
on the right-hand side (RHS) where $g_{\mu\nu} = \eta_{\mu\nu} + \varphi_{\mu\nu}$, we find that the propagator of $\varphi_{\mu\nu}$
in momentum space is schematically given by
\begin{eqnarray}
\frac{1}{p^2 ( p^2 + m^2)} P_{\mu\nu, \rho\sigma}^{(2)} 
\propto \left( \frac{1}{p^2} - \frac{1}{p^2 + m^2} \right) P_{\mu\nu, \rho\sigma}^{(2)},
\label{QG-prop}  
\end{eqnarray}
where $P_{\mu\nu, \rho\sigma}^{(2)}$ describes the projection operator to the spin-$2$ tensor sector. 
This expression shows that quadratic gravity possesses two spin-$2$ degrees of freedom, one of which is a massless spin-$2$ particle
with positive norm, which can be identified with the graviton, and the other is a massive spin-$2$ particle with negative norm,
which is the massive ghost and provides us with the very problem of the unitarity violation.

Next note that the pole structure of this propagator is very similar to that of Pauli-Villars regulator, so it is usually concluded that
it is difficult to get rid of the massive ghost without taking a physically unrealistic limit $m^2 \rightarrow \infty$. Thus, quadratic gravity cannot
be regarded as a viable theory of quantum gravity although it possesses very appealing features of the perturbative renormalizability and asymptotic
freedom.
   
However, this conclusion seems to be too hasty to miss some important characteristic features of quadratic gravity. In particular, the above argument 
focuses on only free part of the action and completely ignores the interactions. Moreover, compared with Einstein gravity which is asymptotically
non-free as in quantum electrodynamics (QED), quadratic gravity is asymptotically free as in quantum chromodynamics (QCD).  
Color confinement, i.e., confinement of quarks and gluons, is one of the central issues in QCD so the massive ghost might be
confined by the interactions, thereby the physical S-matrix becoming unitary. In this article, we wish to investigate 
the possibility of the confinement of the massive ghost.  

The paper is organized as follows: In the next section, we present an idea of confinement of massive ghost in quadratic gravity in the covariant
canonical formalism. In Section 3, we construct a superfield formalism on the six-dimensional superspace and derive field equations for
the massive ghost and the bound state which is a BRST partner of the massive ghost. 
In final section, we draw our conclusion.

\section{Confinement of massive ghost}

As mentioned in the previous section, the classical action in which we are interested takes the form\footnote{We follow the notation and conventions of 
Misner-Thorne-Wheeler (MTW) textbook \cite{MTW}. }:
\begin{eqnarray}
S = \int d^4 x \, \sqrt{-g} \left( \frac{1}{2 \kappa^2} R - \alpha_C C_{\mu\nu\rho\sigma}^2 + \alpha_R R^2 \right),
\label{QG-action}  
\end{eqnarray}
where $\kappa^2 \equiv 8 \pi G$ with $G$ being the Newton constant, $R$ is the scalar curvature and $C_{\mu\nu\rho\sigma}$ is 
the conformal tensor defined by
\begin{eqnarray}
C_{\mu\nu\rho\sigma} &=& R_{\mu\nu\rho\sigma} - \frac{1}{2} ( g_{\mu\rho} R_{\nu\sigma}
- g_{\mu\sigma} R_{\nu\rho} - g_{\nu\rho} R_{\mu\sigma} + g_{\nu\sigma} R_{\mu\rho} )
\nonumber\\
&+& \frac{1}{6} ( g_{\mu\rho} g_{\nu\sigma} - g_{\mu\sigma} g_{\nu\rho} ) R.
\label{C-tensor}  
\end{eqnarray}
Moreover, the coupling constants $\alpha_C$ and $\alpha_R$ are positive and dimensionless. Also recall that these coupling constants
are asymptotically free \cite{Mario, Fradkin, Avramidi}.

Starting with this action, we have already performed a covariant canonical quantization in the de Donder (harmonic) gauge and clarified 
its physical content  \cite{Oda-Can}.\footnote{The covariant canonical quantization of various gravitational theories has also been done 
in the de Donder (harmonic) gauge \cite{Oda-f, Oda-Q, Oda-W, Oda-Saake, Oda-Corfu, Oda-Ohta, Oda-Conf}.}
When we expand the metric in the weak field approximation as\footnote{From now on, indices are raised and lowered with the flat
Minkowski metric $\eta_{\mu\nu}$.}
\begin{eqnarray}
g_{\mu\nu} = \eta_{\mu\nu} + \varphi_{\mu\nu},
\label{eta-exp}  
\end{eqnarray}
with being $| \varphi_{\mu\nu} | \ll 1$, we have found that the metric fluctuation $\varphi_{\mu\nu}$ can be decomposed into \cite{Oda-Can}
\begin{eqnarray}
\varphi_{\mu\nu} = h_{\mu\nu} + 2 \kappa^2 \gamma \psi_{\mu\nu} - \frac{2 (\beta_1 + \beta_2) \kappa^2 \gamma}{3 \beta_1}
\left( \eta_{\mu\nu} + \frac{2}{m^2} \partial_\mu \partial_\nu \right) \phi,
\label{varphi-dec}  
\end{eqnarray}
where $h_{\mu\nu}, \psi_{\mu\nu}$ and $\phi$, respectively denote the graviton, the massive ghost and the massive scalar, 
and they satisfy the following equations:
\begin{eqnarray}
\Box^2 h_{\mu\nu} &=& \partial^\mu h_{\mu\nu} - \frac{1}{2} \partial_\nu h = 0,
\nonumber\\
( \Box - M^2 ) \psi_{\mu\nu} &=& \partial^\mu \psi_{\mu\nu} = \eta^{\mu\nu} \psi_{\mu\nu} = 0,
\nonumber\\
( \Box - m^2 ) \phi &=& 0.
\label{3-eqs}  
\end{eqnarray}
Here $\gamma, \beta_1$ etc. are certain constants \cite{Oda-Can}. The concrete expressions for $h_{\mu\nu}, \psi_{\mu\nu}$ and $\phi$
can be also found in Ref. \cite{Oda-Can}.
   
Next let us focus on the massive ghost $\psi_{\mu\nu}$. Under the BRST transformation corresponding to the infinitesimal general coordinate 
transformation (GCT), the metric $g_{\mu\nu}$ transforms as
\begin{eqnarray}
\delta_B g_{\mu\nu} &\equiv& [ i Q_B, g_{\mu\nu} ] 
\nonumber\\
&=& - ( \nabla_\mu c_\nu + \nabla_\nu c_\mu )
\nonumber\\
&=& - ( c^\alpha \partial_\alpha g_{\mu\nu} + \partial_\mu c^\alpha g_{\alpha\nu} + \partial_\nu c^\alpha g_{\alpha\mu} ),
\label{metric-BRST}  
\end{eqnarray}
where $Q_B$ is the BRST charge and $c_\mu$ is the FP ghost. At the linearized level, it is easy to see that 
the BRST transformation for $\psi_{\mu\nu}$ reads
\begin{eqnarray}
\delta_B \psi_{\mu\nu} \equiv [ i Q_B, \psi_{\mu\nu} ] = - ( c^\alpha \partial_\alpha \psi_{\mu\nu} + \partial_\mu c^\alpha \psi_{\alpha\nu} 
+ \partial_\nu c^\alpha \psi_{\alpha\mu} ) \equiv \Gamma_{\mu\nu}.
\label{psi-BRST}  
\end{eqnarray}
 
The key assumption for confinement of massive ghost is to assume that the composite Heisenberg operator $\Gamma_{\mu\nu}$ has 
an asymptotic field, say $\gamma_{\mu\nu}$, corresponding to a bound state by an interaction between the massive ghost and 
the FP ghost.\footnote{The pioneering work for color confinement along the same idea has been performed in Ref. \cite{Kugo}.}
Of course, in the standard approach of the BRST formalism, we do not assume the existence of 
such a bound state, and then we find that the massive ghost $\psi_{\mu\nu}$ is a physical massive state of spin-2 with negative norm 
which violates the unitarity. However, the assumption of the bound state changes the result radically as will be seen shortly. 

We will make use of the anti-BRST transformation \cite{Curci, Ojima} as well as the BRST transformation since in the next section we work with the
superfield formalism on the six-dimensional superspace $(x^\mu, \theta, \bar \theta)$ \cite{Bonora-Tonin} where both the BRST and anti-BRST 
play a role. The anti-BRST transformation, which is denoted as $\bar \delta_B$ and its anti-BRST charge is $\bar Q_B$, for the massive ghost 
$\psi_{\mu\nu}$ is obtained by replacing the FP ghost $c_\mu$ with the FP anti-ghost $\bar c_\mu$:
\begin{eqnarray}
\bar \delta_B \psi_{\mu\nu} \equiv [ i \bar Q_B, \psi_{\mu\nu} ] = - ( \bar c^\alpha \partial_\alpha \psi_{\mu\nu} + \partial_\mu \bar c^\alpha \psi_{\alpha\nu} 
+ \partial_\nu \bar c^\alpha \psi_{\alpha\mu} ) \equiv \bar \Gamma_{\mu\nu}.
\label{psi-anti-BRST}  
\end{eqnarray}
By the assumption that $\gamma_{\mu\nu}$ is a bound-state asymptotic field of $\Gamma_{\mu\nu}$, $\gamma_{\mu\nu}$ should produce
a pole in a two-point function $\langle 0 | T \, \Gamma_{\mu\nu}(x) \bar \Gamma_{\rho\sigma}(y) | 0 \rangle$ where $\bar \Gamma_{\mu\nu}$ 
is defined in (\ref{psi-anti-BRST}) and is the FP-conjugate operator of $\Gamma_{\mu\nu}$.

Owing to $\{Q_B, \bar Q_B\} = 0$, taking the BRST transformation of $\bar \Gamma_{\mu\nu}$ or the anti-BRST transformation of $\Gamma_{\mu\nu}$  
produces
\begin{eqnarray}
\{ i Q_B, \bar \Gamma_{\mu\nu} \} = - \{ i \bar Q_B, \Gamma_{\mu\nu} \} \equiv B_{\mu\nu},
\label{B-BS}  
\end{eqnarray}
where $B_{\mu\nu}$ is defined by
\begin{eqnarray}
B_{\mu\nu} &=& - ( i B^\alpha \partial_\alpha \psi_{\mu\nu} - \bar c^\alpha \partial_\alpha \Gamma_{\mu\nu} + i \partial_\mu B^\alpha \psi_{\alpha\nu} 
- \partial_\mu \bar c^\alpha \Gamma_{\alpha\nu} 
\nonumber\\
&+& i \partial_\nu B^\alpha \psi_{\mu\alpha} - \partial_\nu \bar c^\alpha \Gamma_{\mu\alpha} )
\nonumber\\
&=& i \bar B^\alpha \partial_\alpha \psi_{\mu\nu} - c^\alpha \partial_\alpha \bar \Gamma_{\mu\nu} + i \partial_\mu \bar B^\alpha \psi_{\alpha\nu} 
- \partial_\mu c^\alpha \bar \Gamma_{\alpha\nu} 
\nonumber\\
&+& i \partial_\nu \bar B^\alpha \psi_{\mu\alpha} - \partial_\nu c^\alpha \bar \Gamma_{\mu\alpha},
\label{B}  
\end{eqnarray}
In deriving this equation, we have used that $\delta_B \bar c^\mu = i B^\mu$ and $\bar \delta_B c^\mu = i \bar B^\mu$ where $B^\mu$ and $\bar B^\mu$ 
are the auxiliary fields, and the relation in the second equality:
\begin{eqnarray}
B^\mu + \bar B^\mu - i ( c^\nu \partial_\nu \bar c^\mu + \bar c^\nu \partial_\nu c^\mu ) = 0.
\label{B-bar-B}  
\end{eqnarray}
Provided that operators $\Gamma_{\mu\nu}, \bar \Gamma_{\mu\nu}$ and $B_{\mu\nu}$ develop bound states with asymptotic fields   
$\gamma_{\mu\nu}, \bar \gamma_{\mu\nu}$ and $\beta_{\mu\nu}$, respectively, then $\gamma_{\mu\nu}, \bar \gamma_{\mu\nu}, \beta_{\mu\nu}$ 
and $\psi_{\mu\nu}$ form a BRST quartet and appear in the physical subspace only in the zero-norm combinations, which is nothing but
confinement of the massive ghost.  

In this way, we can restore the unitarity of the physical S-matrix by making the asymptotic field of the massive ghost
confine in the unphysical Hilbert space. Of course, one of the most important question about  this confinement mechanism is to show that 
the bound states are in fact constructed by the gravitational interaction. For this aim, we would need some non-perturbative approach such as
summing up all radiative corrections in the ladder approximation \cite{Oda0, Oda1}.

\section{Superfield formalism}

In the previous section, we have presented a mechanism for confinement of massive ghost on the basis of the BRST and anti-BRST transformations.
A natural framework where both the BRST and anti-BRST transformations are described in a geometrical way is the superfield formalism on the 
six-dimensional superspace by Bonora and Tonin \cite{Bonora-Tonin}, so we would like to apply this formalism to the present problem of confinement
of the massive ghost.  In the case of QCD, the superfield formalism has also been employed for confinement of gluons \cite{Bonora-Pasti-Tonin}.

Before doing that, let us point out two important remarks on the present mechanism. One of them is that if the bound states described in the previous section
exist, the general coordinate symmetry no longer ensures its BRST and anti-BRST invariances since we have modified them only for the massive
ghost by hand. The other remark is that the classical action (\ref{QG-action}), or more precisely its linearised action breaks the BRST and anti-BRST invariances.      
Since it is not the general coordinate invariance but the BRST and anti-BRST invariances that we have to respect in taking account of the quantum field theories,
we shall reconsider an effective action for the asymptotic fields, $(\psi_{\mu\nu}, \gamma_{\mu\nu}, \bar \gamma_{\mu\nu}, \beta_{\mu\nu})$ 
and understand its physical implications within the framework of the superfield formalism. 

At this stage, let us briefly review the superfield formalism \cite{Bonora-Tonin}.\footnote{Our notation is slightly different from that of \cite{Bonora-Tonin}.}
A generic superfield $\Phi ( x, \theta, \bar \theta )$ is defined on the six-dimensional superspace coordinates $(x^\mu, \theta, \bar \theta)$ 
 where $\theta, \bar \theta$ are two Grassmann coordinates satisfying $\theta^* = - \theta, \, \bar \theta^* = - \bar \theta, \, \theta^2 
 = \bar \theta^2 = \{ \theta, \bar \theta \} = 0$ and anticommuting with FP ghosts $c_\mu$ and $\bar c_\mu$.  Because of Grassmann nature of 
 $\theta$ and $\bar \theta$, the superfield can be expanded into a finite Taylor series as
\begin{eqnarray}
\Phi ( x, \theta, \bar \theta ) = \left. \Phi \right|_0 + \theta \left. \frac{\partial \Phi}{\partial \theta} \right|_0 
+ \bar \theta \left. \frac{\partial \Phi}{\partial \bar \theta} \right|_0 + \bar \theta \theta \left. \frac{\partial^2 \Phi}{\partial \theta \partial \bar \theta} \right|_0,
\label{Superfield}  
\end{eqnarray}
where $\left. {} \right|_0$ denotes setting $\theta = \bar \theta = 0$. Since $\frac{\partial}{\partial \theta}$ and $\frac{\partial}{\partial \bar \theta}$
correspond to the BRST transformation $\delta_B$ and the anti-BRST transformation $\bar \delta_B$, respectively, Eq. (\ref{Superfield}) can be
rewritten as 
\begin{eqnarray}
\Phi ( x, \theta, \bar \theta ) = \Phi (x) + \theta \delta_B \Phi (x) + \bar \theta \bar \delta_B \Phi (x) + \bar \theta \theta \delta_B \bar \delta_B \Phi (x).
\label{Superfield-2}  
\end{eqnarray}
One of the key points in the superfield formalism is that superfields are closed under multiplications, i.e., the product of two superfields becomes
a superfield again. The other is that the coefficient in front of $\bar \theta \theta$ in the superfield is invariant under both BRST and anti-BRST
transformations, so it is a physical observable. To put it differently, taking the partial derivative $\frac{\partial^2}{\partial \theta \partial \bar \theta}$
of the superfield provides us with invariant quantities under the BRST and anti-BRST transformations.

We are now ready to present our superfield formalism of massive ghost.
As for the asymptotic fields of the quartet including the massive ghost $\psi_{\mu\nu}$, let us make a superfield defined as
\begin{eqnarray}
\Phi_{\mu\nu}^{as} (z) \equiv \Phi_{\mu\nu}^{as} ( x, \theta, \bar \theta ) = \psi_{\mu\nu} (x) + \theta \gamma_{\mu\nu} (x) + \bar \theta \bar \gamma_{\mu\nu} (x)
+ \bar \theta \theta \beta_{\mu\nu} (x),
\label{Asym-superfield}  
\end{eqnarray}
where all the component fields on the RHS are functions of space-time coordinates $x^\mu$ and obey the transverse and traceless conditions, e.g.,
$\partial^\mu \gamma_{\mu\nu} = \eta^{\mu\nu} \gamma_{\mu\nu} = 0$. Note that this superfield has the same structure as Eq. (\ref{Superfield-2}). 
Furthermore, for simplicity we have ignored the renormalization constants in front of the component fields on the RHS so that our asymptotic
fields are not properly normalized.   

Next, using the superfield (\ref{Asym-superfield}), let us attempt to construct an effective Lagrangian for these asymptotic fields. The effective
Lagrangian should satisfy the following requirements: First, it must be invariant under the BRST and anti-BRST transformations. Note that this
requirement is automatically satisfied by taking the derivative $\frac{\partial^2}{\partial \theta \partial \bar \theta}$ of a superfield, which is one of
advantages in the superfield formalism at hand. Secondly, the effective Lagrangian must be quadratic in the asymptotic fields since they are
free and non-interacting fields. Thirdly, it must contain the second derivative at most to avoid an additional ghost. Incidentally, even if  this requirement
is imposed on the superfields in the effective Lagrangian, there might appear higher-order derivatives in the component fields after
the functional integrations. 

It is of interest that these three requirements almost fix the form of the effective Lagrangian, which is concretely given by
\begin{eqnarray}
{\cal{L}}_{eff} = \frac{\partial^2}{\partial \theta \partial \bar \theta} \Bigg( \frac{1}{4} \Phi_{\rho\mu\nu}^{as} \Phi^{as \rho\mu\nu}  
- \frac{\zeta}{2} \Phi_{\theta\mu\nu}^{as} \Phi_{\bar \theta}^{as \mu\nu} + \frac{\xi}{2} \Phi_{\mu\nu}^{as} \Phi^{as \mu\nu} \Bigg),
\label{Eff-Lag}  
\end{eqnarray}
where $\zeta$ and $\xi$ are constants. Here we have defined
\begin{eqnarray}
\Phi_{\rho\mu\nu}^{as} &\equiv& \partial_\rho \Phi^{as}_{\mu\nu} = \partial_\rho \psi_{\mu\nu} + \theta \partial_\rho \gamma_{\mu\nu} 
+ \bar \theta \partial_\rho \bar \gamma_{\mu\nu} + \bar \theta \theta \partial_\rho \beta_{\mu\nu},
\nonumber\\
\Phi_{\theta\mu\nu}^{as} &\equiv& \frac{\partial}{\partial \theta} \Phi_{\mu\nu}^{as} = \gamma_{\mu\nu} - \bar \theta \beta_{\mu\nu},
\nonumber\\
\Phi_{\bar \theta\mu\nu}^{as} &\equiv& \frac{\partial}{\partial \bar \theta} \Phi_{\mu\nu}^{as} = \bar \gamma_{\mu\nu} + \theta \beta_{\mu\nu}.
\label{Def-Phi}  
\end{eqnarray}

After integrating by parts, this effective Lagrangian can be cast to the following form in terms of the component fields:
\begin{eqnarray}
{\cal{L}}_{eff} = - \frac{1}{2} \psi_{\mu\nu} ( \Box - 2 \xi ) \beta^{\mu\nu} + \frac{1}{2} \bar \gamma_{\mu\nu} ( \Box - 2 \xi ) \gamma^{\mu\nu}  
+ \frac{1}{2} \zeta \beta_{\mu\nu} \beta^{\mu\nu}.
\label{Eff-Lag2}  
\end{eqnarray}
Actually, it is easy to show explicitly that this effective Lagrangian is invariant under both BRST and anti-BRST transformations.  
Since ${\cal{L}}_{eff}$ is quadratic in $\beta_{\mu\nu}$, one can perform the path integration over $\beta_{\mu\nu}$ whose result reads
\begin{eqnarray}
{\cal{L}}_{eff} = - \frac{1}{8 \zeta} \psi_{\mu\nu} ( \Box - 2 \xi )^2 \psi^{\mu\nu} + \frac{1}{2} \bar \gamma_{\mu\nu} ( \Box - 2 \xi ) \gamma^{\mu\nu}.
\label{Eff-Lag3}  
\end{eqnarray}
Then, field equations for the massive ghost $\psi_{\mu\nu}$ and its BRST partner $\gamma_{\mu\nu}$ can easily be derived to
\begin{eqnarray}
( \Box - 2 \xi )^2 \psi_{\mu\nu} = 0, \qquad
( \Box - 2 \xi ) \gamma_{\mu\nu} = 0.
\label{Dipole-eq}  
\end{eqnarray}
It is obvious that the asymptotic field for the massive ghost is a massive dipole while that for the bound state obeys a massive
Klein-Gordon equation. It is worthwhile to notice that the massive dipole field in general contains a new ghost field but this ghost field 
plays no role in the present theory since it appears in the physical Hilbert space in the zero-norm combinations owing to the BRST
invariance.

We would like to understand the physical implications of the effective Lagrangian and the field equations for the asymptotic fields. 
First of all, it is remarkable that comparison between Eq. (\ref{3-eqs}) and Eq. (\ref{Dipole-eq}) reveals that the massive ghost $\psi_{\mu\nu}$ 
has become from a massive field to a massive dipole field because of an appearance of the bound state $\Gamma_{\mu\nu}$.  Actually,
the terms involving $\psi_{\mu\nu}$ and $\gamma_{\mu\nu}$ in the effective Lagrangian ${\cal{L}}_{eff}$ of Eq. (\ref{Eff-Lag2}) has the same structure
as the Froissart model \cite{Froissart}. The field equations for $\beta_{\mu\nu}$ and  $\psi_{\mu\nu}$ are respectively given by
\begin{eqnarray}
( \Box - 2 \xi ) \psi_{\mu\nu} = 2 \zeta \beta_{\mu\nu}, \qquad
( \Box - 2 \xi ) \beta_{\mu\nu} = 0.
\label{Field-eqs}  
\end{eqnarray}
From these field equations, we can express $\psi_{\mu\nu} (x)$ in terms of $\beta_{\mu\nu} (x)$ as
\begin{eqnarray}
\psi_{\mu\nu} = \tilde \psi_{\mu\nu} + \frac{\zeta}{2 \xi} ( x^\rho \partial_\rho +c ) \beta_{\mu\nu},
\label{psi-sol}  
\end{eqnarray}
where $\tilde \psi_{\mu\nu} (x)$ obeys the equation
\begin{eqnarray}
( \Box - 2 \xi ) \tilde \psi_{\mu\nu} = 0,
\label{tilde-psi}  
\end{eqnarray}
and $c$ is an arbitrary constant. Changing the value of the constant $c$ is equivalent to adding $\tilde \psi_{\mu\nu}$, so we can take any $c$, for instance
$c = 1$. From these equations, we find that the massive dipole field is constructed out of two massive fields. In other words,
an emergence of the bound state, thereby the massive ghost being confined, is achieved by an acquisition of a new dynamical degree of freedom 
of the massive ghost. It is then natural to ask ourselves whether a similar mechanism occurs in QCD or not. Indeed, in QCD it is suggested 
that color confinement is realized only when massless gluons become massive by absorbing a new degree of freedom \cite{Chaichian}.  

Secondly, an important question in the present idea is relevant to the problem of how we prove that there is a bound state which has precisely the same magnitude 
of the mass as that of the massive ghost. In the classical action (\ref{QG-action}), there is only the (reduced) Planck mass scale so it is natural to set $2 \xi = M_{Pl}^2  
\equiv \frac{1}{\kappa^2}$.  However, it is expected that various interactions with the gravitational fields and matter fields change the magnitude 
of the mass of both the massive ghost and its bound state separately by radiative corrections, so it is nontrivial to show that they have the same 
magnitude of the mass. In our formalism, as symmetries only the BRST and anti-BRST symmetries play a role, so it is these symmetries that
gurantee the existence of the same magnitude of the mass of the massive ghost and its bound state. Thus, without worrying this issue,
the remaining question is to show that there is really the unique bound state in the BRST transformation of the massive ghost. Then, we hope that
the method developed in Refs. \cite{Oda0, Oda1} provides us with a useful tool for clarifying this question.

Finally, let us comment on the relation between the present formalism of the massive ghost in quadratic gravity and that of gluons in QCD \cite {Bonora-Pasti-Tonin}. 
In the latter formalism, it is shown that the quartet mechanism of color confinement by Kugo \cite{Kugo} leads to a violation of the cluster property
on the basis of the superfield formalism. Their conclusion comes from the observation that the propagator of the gluon field in momentum space becomes the massless 
dipole propagator given by
\begin{eqnarray}
\Delta_{\mu\nu} (p) = \frac{1}{p^4} \left( \eta_{\mu\nu} - \frac{p_\mu p_\nu}{p^2} \right).
\label{Gluon-prop}  
\end{eqnarray}
Here note that the massless dipole propagator yields a linear potential $V(r) \propto r$ at the static case. In the scenario of the quark confinement,
a quark and an anti-quark might be connected by a stringlike object which produces the linear potential to confine the pair of the quark and anti-quark into a hadron,
which means the failure of the cluster property \cite{Araki, Strocchi}. However, in the formalism in Ref. \cite {Bonora-Pasti-Tonin}
as well, it is natural to add to their effective Lagrangian the last term on the RHS of (\ref{Eff-Lag}), which we call ``mass term''.
Then, we have the massive dipole propagator for the gauge field $A_\mu^a (x)$ instead of the massless one. This situation coincides with the
result by Chaichian and Nishijima in \cite{Chaichian}.  The detail of this study will be reported in a separate publication.

\section{Conclusions}

In this article, we have studied the problem of confinement of massive ghost which violates the unitarity of the physical S-matrix in quadratic gravity. 
Using the BRST and anti-BRST transformations, we have shown that if there is a bound state in the channel of the massive ghost and Faddeev-Popov ghost 
the massive ghost is confined in the zero-norm states through the BRST quartet mechanism, thereby the unitarity being restored. This confinement
mechanism of the massive ghost was previously proposed by means of only the BRST transformation \cite{Kawasaki, Kimura2}, but our presentation 
in terms of the both BRST and anti-BRST transformations provides a more unified picture of this mechanism since our formalism can be nicely formulated 
in the superfield formalism by Bonora and Tonin \cite{Mario}.

Furthermore, on the basis of the superfield formalism respecting the BRST and anti-BRST transformations, it has been shown that the asymptotic field 
of the massive ghost must be a massive dipole whereas that of the bound state obeys a massive Klein-Gordon equation. This situation may be 
of some similarity to color confinement in quantum chromodynamics (QCD) where it is conjectured that not a massless but a massive gluon is 
in fact confined \cite{Chaichian}.

The remaining important problem is to show in an explicit manner that the BRST transformation of the massive ghost constitues a bound
state. For this purpose, it is necessary to make use of some non-perturbative method to deal with the strong coupling physics.
We hope that the method, which was developed in terms of the ladder approximation \cite{Oda0, Oda1}, plays a role.


\end{document}